\documentclass[letter,twocolumn]{jpsj2} 

\def\runauthor{Online-Journal Subcommittee}

\title{
Anomalous Metal-Insulator Transition in Filled Skutterudite CeOs$_4$Sb$_{12}$
}

\author{%
Yoshiki {\sc Imai}, 
Keitaro {\sc Sakurazawa} and Tetsuro {\sc Saso}
}

\inst{%
Department of Physics, Saitama University, Saitama, 338-8570
}

\recdate{\hspace{3cm}}

\abst{
Anomalous metal-insulator transition observed in filled skutterudite CeOs$_4$Sb$_{12}$ is investigated by constructing the effective tight-binding model with the Coulomb repulsion between f electrons.   By using the mean field approximation, magnetic susceptibilities are calculated and the phase diagram is obtained.
When the band structure has a semimetallic character 
with small electron and hole pockets at $\Gamma$ and H points, a spin density wave transition with the ordering vector $\mathbf{Q}=(1,0,0)$  occurs due to the nesting property of the Fermi surfaces. Magnetic field enhances this phase in accord with the experiments.
}

\kword{filled-skutterudite compounds, CeOs$_4$Sb$_{12}$, field induced metal-insulator transition
}
\begin{document}
\maketitle
Filled-skutterudite compounds RM$_4$X$_{12}$ (R: rare earth, M: Fe, Ru, and Os, and X: P, As, and Sb) with the unique body-centered cubic structures have attracted much interest due to wide variety of physical phenomena, e.g. magnetic and quadrupole ordering, Kondo insulating behavior, unconventional superconductivity, etc.\cite{baue02,suga05}

Recently, anomalous phase transition from paramagnetic metal (PM) to spin density wave (SDW) state was found  at $T=$0.8K in CeOs$_4$Sb$_{12}$. Surprisingly, the phase boundary is shifted to higher temperatures by applying the external magnetic field \cite{nami03,suga05}.   Such increase of the transition temperature have been often observed in the systems with quadrupole-ordering,\cite{quadrupole} but the magnetic measurement\cite{baue01} suggests that the crystal-field (CF) ground state of f electron is $\Gamma_7$ which has no quadrupole moment, and the specific heat measurement\cite{suga05} shows only a small entropy change at the transition, excluding local moment ordering but suggesting an SDW or CDW transitions with an opening of a gap over a small portion of the Fermi surfaces of itinerant electronic states.
The neutron scattering experiment\cite{yang05} was done but the CF was not clearly observed.
The NQR measurements\cite{yogi05} claims the first-order transition since the hysterisis was observed, but the specific heat data\cite{suga05} clearly show peaks at the transition indicating the second order transition with some short range order.

CeOs$_4$Sb$_{12}$ shows semiconducting behavior in the resistivity experiment with a small charge gap ($E_{\rm g} \sim 10$K) at low temperatures.\cite{baue01} The overall structure of the optical conductivity and its temperature dependence show similar behaviors to those of CeRu$_4$Sb$_{12}$ which exhibits a pseudo-gap and the mid-infrared peak with strong temperature-dependence.\cite{matu03,Dordevic_01} 
Although the band structure of CeOs$_4$Sb$_{12}$ obtained by the band calculation is quite similar to that of CeRu$_4$Sb$_{12}$\cite{hari03}, CeOs$_4$Sb$_{12}$ shows semiconducting behavior, whereas CeRu$_4$Sb$_{12}$ is metallic in experiment \cite{baue01}.  

In this letter, we construct the effective model with the characteristic feature of the filled skutterudite compound CeOs$_4$Sb$_{12}$ and investigate the anomalous metal-insulator transition observed in this material.  

The basic crystal structure of filled skutterudites is that the X$_{12}$ clusters, filled by R in the center, form the body-centered cubic (BCC) lattice.  The f orbitals of rare-earth atoms strongly hybridize with the p orbitals in X$_{12}$ clusters close the Fermi level.  
The X$_{12}$ cluster has the T$_{\rm h}$ symmetry, but its effect on $J$=5/2 f$^1$ state in Ce is the same as O$_{\rm h}$.

In order to investigate the properties of Ce-skutterudites, we previously proposed a simplified tight-binding band model for CeRu$_4$Sb$_{12}$ consisting of the single valence band and the seven f states with the spin-orbit interaction and under the cubic crystalline field.\cite{muto04}   The former is represented by the top-most one of the valence bands made up of the p-orbitals on X$_{12}$ clusters, and has the A$_{u}$ symmetry which hybridizes with the f-state in the cluster center.   The Coulomb interaction was introduced to all the band equally to  make the use of the dynamical mean-field approximation possible.

Here, we simplify the model further to investigate the magnetic properties of CeOs$_4$Sb$_{12}$.  The present model consists of a single f band (representing the $J$=5/2, $\Gamma_7$ ground states) and a single valence band (the above-mentioned A$_{u}$ band), hybridizing with each other.
Then Hamiltonian is given by
\begin{eqnarray}
H&=&\sum_{{\mathbf k},\sigma}\epsilon^{c}_{\mathbf k}
c^{\dag}_{{\mathbf k},\sigma}c_{{\mathbf k},\sigma}
+\sum_{{\mathbf k},\sigma}\epsilon^{f}_{\mathbf k}
f^{\dag}_{{\mathbf k},\sigma}f_{{\mathbf k},\sigma}\nonumber \\
&+&\sum_{{\mathbf k},\sigma}
\bigg(Vc^{\dag}_{{\mathbf k},\sigma}
f_{{\mathbf k},\sigma}+h.c\bigg)
+U\sum_{}n^{f}_{i\uparrow}n^{f}_{i\downarrow}\nonumber \\
&+&B_{\rm ext}\sum_{{\mathbf k},\sigma}
\sigma \bigg( n^{c}_{{\mathbf k},\sigma}
             +\beta n^{f}_{{\mathbf k},\sigma} \bigg)
\label{ham}
\end{eqnarray}
where $c^{\dag}_{{\mathbf k},\sigma}$($f^{\dag}_{{\mathbf k},\sigma}$) creates  a conduction (f) electron with the momentum ${\mathbf k}$ and pseudo-spin $\sigma$ (representing $\Gamma_7$ doublet), and $n^{f}_{i\sigma}$=$f^{\dag}_{i,\sigma}f_{i,\sigma}$ with the site index $i$.  We have assumed the BCC tight-binding band with nearest and next-nearest hopping for conduction band, $\epsilon^{c}_{\mathbf k}=\alpha_{\rm c} t_{\mathbf k}$,
where 
$t_{\mathbf k}=\cos(k_{x}a/2)\cos(k_{y}a/2)\cos(k_{z}a/2)+\alpha_{2}[\cos(k_{x}a)+\cos(k_{y}a)+\cos(k_{z}a)]$.  
The lattice constant $a$ is set equal to unity in the following.
$\alpha_{\rm c}$ and $\alpha_2$ determine the hopping amplitude.
The f state is assumed to have the dispersion $\epsilon^{f}_{\mathbf k}=E_{f} + \alpha_{\rm f}t_{\mathbf k}$, namely the dispersive part is proportional to $\epsilon^{c}_{\mathbf k}$ but with a small coefficient $\alpha_{\rm f}$.
This would be a reasonable choice since both X$_{12}$ and R occupy the same BCC sites.
$V$ is the hybridization between conduction and f electrons.  Its ${\mathbf k}$-dependence is neglected.  These choice of our model band immensely facilitates the calculations for the strong correlation, which will be reported in a separate paper.\cite{saso05}  Actually, of course, the f-dispersion is created through the X$_{12}$ clusters, so that it must be self-consistently determined with $\epsilon^{c}_{\mathbf k}$ and the hybridization $V_{\mathbf k}$.
$U$ is the Coulomb repulsion between f electrons.  
It is well-known that $\epsilon^{c}_{\mathbf k}$ and $\epsilon^{f}_{\mathbf k}$ have perfect nesting characters for $\alpha_2=0$, $V$=0, and at half-filling.
The last term in eq.  (\ref{ham}) represents the Zeeman energy, where $B_{\rm ext}$ denotes the external magnetic field along the z direction and $\beta$ is coefficient of effective magnetic moment, which corresponds to $g_{J}\langle J^{f}_{z} \rangle$=6/7$\times$5/6=5/7 for Ce.  

Without the Coulomb interaction, the diagonalized energy dispersions are obtained as $E_{\mathbf k}^{\pm}=[\epsilon^{c}_{\mathbf k}+\epsilon^{f}_{\mathbf k}\pm \{(\epsilon^{c}_{\mathbf k}-\epsilon^{f}_{\mathbf k})^{2}+4V^{2}\}^{1/2}]/2$, which are the so-called hybridized bands, but since $\epsilon^{f}_{\mathbf k}$ has also a finite dispersion, the two bands can have finite overlap if $\alpha_{\rm f}$ is not too small. 

\begin{figure}[tb]
\begin{center}
\includegraphics[width=7cm]{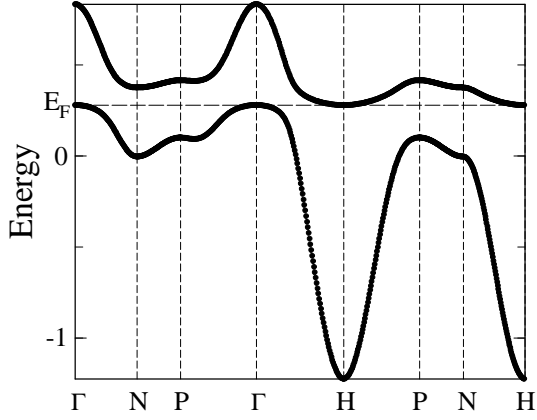}
\end{center}
\caption{
The band structure for the non-interacting system without the external field 
where $\alpha_{\rm c}$=1.0, $\alpha_{\rm f}$=0.03, $\alpha_{2}$=0.07, $E_{\rm f}$=0.3, and $V$=0.15. 
${\rm E_{F}}$ represents the Fermi level. }
\label{band}
\end{figure}
Considering the empirical trend of the lattice constant and the energy gap of Ce-skutterudites, ${\rm CeOs_{4}Sb_{12}}$ must be a semimetal \cite{suga05}. 
Our assumed band structure is shown in Fig. \ref{band}, which has a semi-metallic character: the top of lower band ($\Gamma$ point) and the bottom of upper band (H point) slightly touch the Fermi level, respectively. The overlapping of the bands is about 0.001$\alpha_{c}$. 

Although our band structure is quite simplified, Fig. \ref{band} captures the low-energy structure of the band calculation for CeOs$_4$Sb$_{12}$\cite{hari03} rather well. 
In the published band calculation,\cite{hari03} one of the band made of $\Gamma_8$ states has a minimum at $\Gamma$ point and almost touches with the valence band there, but it is neglected here. Also, only the single valence band is taken into account and the lower bands are completely neglected.  These may affect the quantitative results but at least the qualitative feature may remain unchanged. The existence of $\Gamma_{8}$ band may be a cause of the lower temperature anomaly \cite{yogi05}. 

In order to investigate a possibility of a magnetic transition and an anomalous magnetic field effect, the random phase approximation (RPA) is employed.  Since RPA overestimates the transition temperature, we use the small value of the Coulomb interactions here, which is considered as the renormalized one at the low-energy regime in the rare-earth compounds. 
Furthermore, since the Coulomb interaction acts only on f electrons, ordered states will be mainly determined by the localized f electrons, so that only the f-f component of the correlation function is considered, for simplicity. It is essential for the following theory to calculate both the parallel and perpendicular elements to the external field of the static susceptibility, which are given by
\begin{eqnarray}
&&\chi^{x} ({\mathbf q})=\chi^{y} ({\mathbf q})=\frac{\beta^{2}}{2}
\frac{\chi^{0}_{\uparrow \downarrow}({\mathbf q})}
{1-U\chi^{0}_{\uparrow \downarrow}({\mathbf q})},\\
&&\chi^{z} ({\mathbf q})=\frac{\beta^{2}}{4}
\frac{
     \chi^{0}_{\uparrow \uparrow}({\mathbf q})
    +\chi^{0}_{\downarrow \downarrow}({\mathbf q})
  +2U\chi^{0}_{\uparrow \uparrow}({\mathbf q})
     \chi^{0}_{\downarrow \downarrow}({\mathbf q})
     }
    {1-U^{2}\chi^{0}_{\uparrow \uparrow}({\mathbf q})
            \chi^{0}_{\downarrow \downarrow}({\mathbf q})},
\end{eqnarray}
where $\chi^{0}_{\sigma \sigma'}$ is the band susceptibility, which is defined by  
\begin{eqnarray}
&&\chi^{0}_{\sigma \sigma'} ({\mathbf q})=
  \frac{-T}{N}
  \sum_{{\mathbf k},\omega_{n}}
    g^{\rm f}_{\sigma}({\mathbf k}+{\mathbf q},{\rm i}\omega_{n})
    g^{\rm f}_{\sigma'}({\mathbf k},{\rm i}\omega_{n}),\\
&&g^{\rm f}_{\sigma}({\mathbf k},{\rm i} \omega_{n})
   =\frac{1}
     { {\rm i}\omega_{n} +\mu -\epsilon^{f}_{{\mathbf k}\sigma}
      -{\displaystyle \frac{V^{2}}{{\rm i}\omega_{n} +\mu -\epsilon^{c}_{{\mathbf k}\sigma}}} }.
\end{eqnarray}
$T$ is the temperature and $N$ is the number of sites. $\mu$ is the chemical potential and $\omega_{n}(=(2n+1)\pi T)$ represents the odd Matsubara frequency. $\epsilon^{c}_{{\mathbf k}\sigma}$ and $\epsilon^{f}_{{\mathbf k}\sigma}$ are the energy dispersion with the Zeeman terms. Note that we use the following relation that $\chi_{\downarrow \uparrow}({\mathbf q})$ is always equal to $\chi_{\uparrow \downarrow}({\mathbf q})$ within the RPA to derive $\chi^{x}$. 

We numerically calculate the above-mentioned equations. The summations are efficiently carried out by using the fast Fourier transform with $N=128\times 128\times 128$ points in the ${\mathbf k}$ summation and 4096 Matsubara frequencies in the $\omega_{n}$ summation. Note that hopping amplitude $\alpha_{\rm c}$ is taken to be unity for simplicity. 

\begin{figure}[tb]
\begin{center}
\includegraphics[width=8cm]{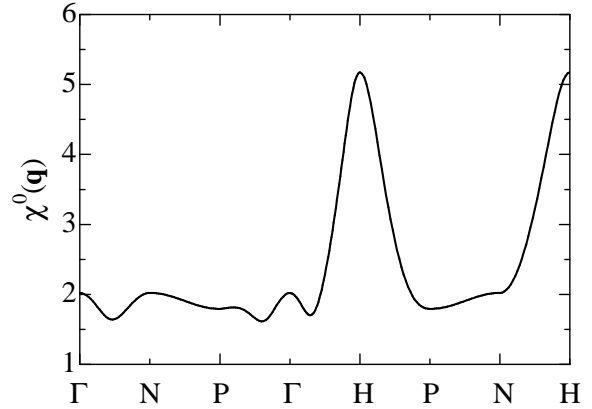}
\end{center}
\caption{
Static irreducible susceptibility for various momenta without the external field at $T=0.002$ where $\chi^{0}({\bf q})=\chi^{0}_{\uparrow \downarrow}({\bf q})=\chi^{0}_{\uparrow \uparrow}({\bf q})=\chi^{0}_{\downarrow \downarrow}({\bf q})$. }
\label{chi_b0}
\end{figure}

To discuss the instability to ordered states, the band susceptibility, $\chi^{0}({\mathbf q}) \equiv \chi^{0}_{\uparrow \uparrow} ({\mathbf q})=\chi^{0}_{\downarrow \downarrow} ({\mathbf q})=\chi^{0}_{\uparrow \downarrow} ({\mathbf q})$ for $B_{\rm ext}$=0, is shown in Fig. \ref{chi_b0}. 
The largest value appears at H point ${\mathbf q}$=$(1,0,0)$ in unit of $2\pi/a$. Since the Fermi surfaces of electrons and holes are quite small spheres, the peak of $\chi^{0}({\mathbf q})$ appears at ${\mathbf q}=(1,0,0)$ at low temperatures, which corresponds to the difference between the top of lower band ($\Gamma$ point) and the bottom of upper band (H point) in Fig. \ref{band}. Recently, the result of the neutron scattering for ${\rm CeOs_{4}Sb_{12}}$ at $B_{\rm ext}=0$ found that the nesting vector corresponds to ${\mathbf q}=(1,0,0)$\cite{iwas05}. Therefore our obtained result of the susceptibilities is consistent with the experiment at $B_{\rm ext}=0$.

The basic lattice structure of ${\rm CeOs_{4}Sb_{12}}$ is BCC. For the single band Hubbard model at half-filling with only the nearest neighbor hopping, there exists the perfect nesting at the wave vector ${\mathbf q}=(1,0,0)$.  In that case, it is known that the infinitesimally small $U$ causes the metal-SDW transition at $T$=0. Although the perfect nesting is lost in the present model due to the hybridization and the next nearest neighbor hoppings, the Fermi surfaces are in the vicinity of the nesting. Therefore, when the external field is absent, the metal-SDW transition occurs at $U_{c}\sim 0.2$ at $T$=0.002 within the RPA in the present model. 

\begin{figure}[tb]
\begin{center}
\includegraphics[width=8cm]{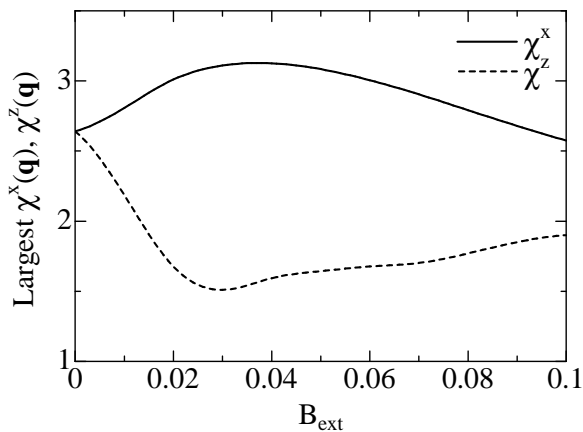}
\end{center}
\caption{The largest susceptibilities as a function of the magnetic field at $T$=0.002 and $U$=0.0. Solid (dashed) line represents $\chi^{x}({\mathbf q})$ ($\chi^{z}({\mathbf q})$). 
 }
\label{max_chi}
\end{figure}

Next, let us discuss effects due to the external magnetic field. The field dependence of the largest values of susceptibilities is shown in Fig. \ref{max_chi}. 
With increasing the magnetic field along the $z$ direction, $\chi^{x}$ is first enhanced at small field region. As the external field is increased further, $\chi^{x}$ is suppressed. The vector ${\mathbf q}$ which gives the largest values of susceptibilities does not change within $B_{\rm ext} < 0.2$. In contrast to the behavior of $\chi^{x}$, $\chi^{z}$ is first suppressed. This behavior is reasonable since when the weak external field is applied, the spin-spin correlation functions along the external field are generally suppressed. Furthermore, with increasing the external field further, maximum value of $\chi^{z}$ is enhanced. In this range of field, the vector ${\mathbf q}$ which gives the maximum of the susceptibilities is gradually changed. Both of susceptibilities are suppressed under larger external field (not shown). 
These behaviors and the mechanisms are similar to our previous theory for CeRhIn$_5$ where more precise FLEX (fluctuation-exchange) method was used.\cite{saku05}

In general, the response functions are strongly affected by the form of Fermi surface. The present band structure at $B_{\rm ext}=0$ is close to the perfect nesting, so that the change of magnetic susceptibilities becomes quite large for the external field. Therefore, it is possible that when the weak external field is applied, the nesting condition is rather improved because one of the spin band is shifted by the field and approaches the nesting condition. 

\begin{figure}[tb]
\begin{center}
\includegraphics[width=8cm]{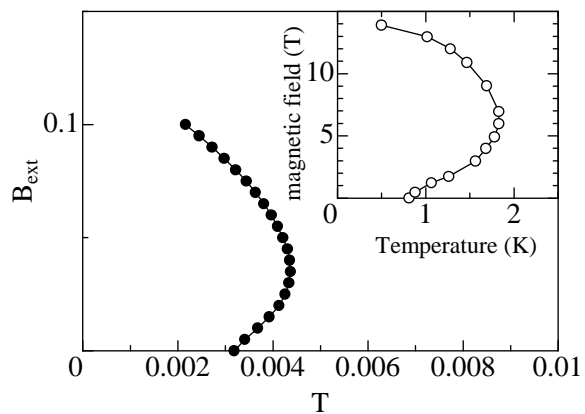}
\end{center}
\caption{The $T$-$B$ Phase diagram of the metal-SDW transition at $U=0.2$. The inset shows the experimental result obtained by Sugawara {\it et al}. \cite{suga05}
 }
\label{phase_diagram}
\end{figure}

These results indicate that the presence of the external field enhances instability for ordered states. In the 3-dimensional system, ordered states can occur at a finite temperature. Namely, the anomalous metal-insulator transition may occur in the present model with appropriate value of the Coulomb repulsion. Figure \ref{phase_diagram} shows the temperature-external field phase diagram of metal-SDW transition. Note that the instability occurs not in $\chi^{z}$(${\mathbf q}$=$(1,0,0)$) but in $\chi^{x}$(${\mathbf q}$=$(1,0,0)$) (and $\chi^{y}$), which indicates the appearance of the magnetic moment component perpendicular to the field. It can be checked by an experiment in a future. 
In the small external field region, the phase transition temperature $T_{\rm N}$ is enhanced with increasing $B_{\rm ext}$, so that the slope of phase boundary becomes positive. Thus the anomalous phase transition observed in ${\rm CeOs_{4}Sb_{12}}$ can be qualitatively reproduced. 

Note here that in order to calculate magnetic susceptibilities, we employ the RPA, which is insufficient to take the effect of quantum fluctuations into account. Quantum fluctuations particularly become important in the vicinity of quantum critical points where antiferromagnetic correlations are enhanced \cite{saku05}. However, correlation effects are essential in the weak external field region. By taking quantum fluctuations into account e.g. via FLEX approximation,\cite{saku05} the metal-SDW transition temperature is reduced and the transition boundary as a whole shifts to the low temperature region. In particular, the shift becomes conspicuous at the weak external field region. However, overall structure of the phase boundary obtained here may not qualitatively changed. 

Furthermore, we comment the multi-band effects of f orbitals. In our model, we neglect the degeneracy of f orbitals, for simplicity. 
However, in the published band calculation, one of the bands made from the 4f $\Gamma_8$ states has a minimum at $\Gamma$ point, though slightly above $E_F$, which is neglected in our model.
However, we consider that the nesting property between the top of the valence band at $\Gamma$ and the bottom of the $\Gamma_7$ band at H is the main source of the SDW transition and anomalous phase diagram.
Therefore qualitative properties of ${\rm CeOs_{4}Sb_{12}}$ can be reproduced by using the present simple model. 

A similar anomalous magnetic field effect is also investigated for the Kondo insulator \cite{mila04,ohas04}, in which the ordered phase appears at finite field since the perpendicular element of spin-spin correlations to the external field is enhanced. 

In summary, we investigate the anomalous phase transition observed in ${\rm CeOs_{4}Sb_{12}}$. By constructing the effective tight-binding model with the Coulomb interaction between f electrons, whose band structure is in the vicinity of the nesting, the metal-SDW phase boundary is determined within the mean field approximation. The nesting condition of Fermi surfaces is improved when the weak external field is applied, so that the slope of metal-SDW phase boundary becomes positive and the anomalous transition can be reproduced. 
A small moment perpendicular to the applied magnetic field should be induced, which can be checked by a future experiment.


We would like to thank Professors M.  Kohgi, K.  Iwasa, H.  Sugawara, H.  Harima, K.  Takegahara and H.  Kontani for useful informations and illuminating discussions.  
This work was supported by The Grant-in-Aid from the Ministry of Education, Science and Culture: ``Evolution of New Quantum Phenomena Realized in the Filled Skutterudite Structure'', No.  16037204.
A part of computations was done at the Supercomputer Center at the Institute for Solid State Physics, University of Tokyo. 

\end{document}